\input harvmac
\input tables
\noblackbox
% Poor man's Blackboard Bold characters often used :
\def\inbar{\,\vrule height1.5ex width.4pt depth0pt}
\def\IB{\relax{\rm I\kern-.18em B}}
\def\IC{\relax\hbox{$\inbar\kern-.3em{\rm C}$}}
\def\ID{\relax{\rm I\kern-.18em D}}
\def\IE{\relax{\rm I\kern-.18em E}}
\def\IF{\relax{\rm I\kern-.18em F}}
\def\IG{\relax\hbox{$\inbar\kern-.3em{\rm G}$}}
\def\IH{\relax{\rm I\kern-.18em H}}
\def\II{\relax{\rm I\kern-.18em I}}
\def\IK{\relax{\rm I\kern-.18em K}}
\def\IL{\relax{\rm I\kern-.18em L}}
\def\IM{\relax{\rm I\kern-.18em M}}
\def\IN{\relax{\rm I\kern-.18em N}}
\def\IO{\relax\hbox{$\inbar\kern-.3em{\rm O}$}}
\def\IP{\relax{\rm I\kern-.18em P}}
\def\IQ{\relax\hbox{$\inbar\kern-.3em{\rm Q}$}}
\def\IR{\relax{\rm I\kern-.18em R}}
\font\cmss=cmss10 \font\cmsss=cmss10 at 7pt
\def\IZ{\relax\ifmmode\mathchoice
{\hbox{\cmss Z\kern-.4em Z}}{\hbox{\cmss Z\kern-.4em Z}}
{\lower.9pt\hbox{\cmsss Z\kern-.4em Z}}
{\lower1.2pt\hbox{\cmsss Z\kern-.4em Z}}\else{\cmss Z\kern-.4em Z}\fi}

%
% Journals
%

\def\NP{{\it Nucl. Phys.\ }}

\def\PL{{\it Phys. Lett.\ }}

\def\CMP{{\it Comm. Math. Phys.\ }}

\def\IJMP{{\it Int. Jour. Mod. Phys.\ }}

%
%Fonts
%
\font\tiau=cmcsc10
%
%References
%
\lref\joep{J. Polchinski, ``Dirichlet-Branes
and Ramond-Ramond Charges,'' hep-th/9510017.}
\lref\andyc{A. Strominger, ``Massless Black Holes and Conifolds in
String Theory,'' hep-th/9504090, {\it Nucl. Phys.} {\bf B451} (1995) 96.}
\lref\pols{J. Polchinski and A. Strominger, ``New Vacua for Type II String
Theory,'' hep-th/9510227.}
\lref\small{E. Witten, ``Small Instantons in String Theory,'' hep-th/9511030.}
\lref\chs{C. Callan, J. Harvey, and A. Strominger, ``Worldbrane Actions
for String Solitons,'' {\it Nucl. Phys.} {\bf B367} (1991) 60.}
\lref\hors{G. Horowitz and A. Strominger, ``Black Strings and p-Branes,''
{\it Nucl. Phys.}{\bf B360} (1991) 197.}
\lref\dlp{J. Dai, R. Leigh, and J. Polchinski, ``New Connections Between String
Theories,'' {\it Mod. Phys. Lett.}{\bf A4} (1989) 2073.}
\lref\joed{J. Polchinski and E. Witten, ``Evidence for Heterotic
Type I String Duality,'' hep-th/9510169.}
\lref\hullt{C. Hull and P. Townsend, ``Unity of Superstring Dualities,''
hep-th/9410167, Nucl. Phys. {\bf B438} (1995) 109.}
\lref\witt{E. Witten,``String Theory Dynamics in Various Dimensions,''
{\it Nucl. Phys.} {\bf B443} (1995) 1995, hep-th/9503124.}
\lref\hs{J. A. Harvey and A. Strominger, ``The Heterotic String is a
Soliton,'' {\it Nucl. Phys. B} {\bf B449} 535 (1995), hep-th/9504047.}
\lref\petr{P. Horava and E. Witten, ``Heterotic and Type I
String Dynamics From Eleven Dimensions,'' hep-th/9510209.}
\lref\jhs{J. Schwarz, ``The Power of M Theory,'' hep-th/9510086.}
\lref\bps{J. Harvey and G. Moore, ``Algebras, BPS states,
and Strings,'' hep-th/9510182.}
\lref\seiberg{N. Seiberg, ``Observations on the Moduli Space of
Superconformal Field Theories," {\it Nucl. Phys. }{\bf B303} (1988) 286.}
\lref\aspmor{P. Aspinwall and D. Morrison, ``String Theory on K3
surfaces," to appear in ``Essays on Mirror Manifolds 2,'' hep-th/9404151.}
\lref\aspinpt{P. Aspinwall, ``Enhanced Gauge Symmetries and K3
Surfaces," hep-th/9507012, \PL {\bf B357} (1995) 329.}
\lref\chpol{S. Chaudhuri and J. Polchinski, ``Moduli Space of CHL Strings,''
hep-th/9506048.}
\lref\chl{S. Chaudhuri, G. Hockney and J. Lykken, ``Maximally Supersymmetric
String Theories in $D<10$,'' {\it Phys. Rev. Lett.} {\bf 75} 2264
(1995), hep-th/9505054.}
\lref\clow{S. Chaudhuri and D.A. Lowe, ``Type IIA - Heterotic Duals with
Maximal Supersymmetry,'' to appear in {\it Nucl. Phys. B}, hep-th/9508144.}
\lref\nikulin{V.V. Nikulin, ``Finite Automorphism Groups of Kahler K3
Surfaces,'' {\it Trans. Moscow Math. Soc.} {\bf 2} (1980) 71.}
\lref\sens{A. Sen, ``String-String Duality Conjecture in Six Dimensions
and Charged Solitonic Strings,'' {\it Nucl. Phys.} {\bf B450}
(1995) 103, hep-th/9504027.}
\lref\kv{S. Kachru and C. Vafa, ``Exact Results for N=2 Compactifications
of Heterotic Strings,'' \NP {\bf B450} (1995) 69, hep-th/9505105.}
\lref\fhsv{S. Ferrara, J. Harvey, A. Strominger and C. Vafa, ``Second
Quantized Mirror Symmetry,'' {\it Phys. Lett.}{\bf B361} (1995) 59,
hep-th/9505162.}
\lref\jeff{A. Dabholkar and J. Harvey, {\it Phys. Rev. Lett.}
{\bf 63} (1989) 478; A. Dabholkar, G. Gibbons, J. Harvey, and
F. Ruiz-Ruiz, \NP {\bf B340} (1990) 33.}
\lref\schsenb{J. Schwarz and A. Sen, ``The Type IIA Dual of the
Six-Dimensional CHL Compactification,'' {\it Phys. Lett.} {\bf B357}
323 (1995), hep-th/9507027.}
\lref\aspin{P. Aspinwall, ``Some Relationships Between Dualities in
String Theory,'' hep-th/9508154.}
\lref\aspe{P. Aspinwall, ``An N=2 pair and a phase transition,''
hep-th/9510142.}
\lref\witstr{E. Witten, ``Some Comments on String Dynamics,''
hep-th/9507121.}
\lref\vafwit{C. Vafa and E. Witten, ``Dual String Pairs with N=1
and N=2 Supersymmetry in Four Dimensions,'' hep-th/9507050.}
\lref\niem{H. Niemeier, {\it Jour. Number Theory} {\bf 5} (1973) 142.
J. Conway and N. Sloane, {\it Jour. Number Theory} {\bf 15} (1982) 83.}
\lref\lsw{W. Lerche, A. N. Schellekens, and N. P. Warner, ``Lattices
and Strings,'' {\it Phys. Rep.} {\bf 177} (1989) 1.}
\lref\narain{K.S. Narain, \PL {\bf 169B} (1986) 41; K.S. Narain,
M.H. Sarmadi and E. Witten, \NP {\bf B279} (1987) 369.}
\lref\hls{J.A. Harvey, D.A. Lowe and A. Strominger, ``N=1 String Duality,''
hep-th/9507168, \PL {\bf B362} (1995) 65.}
\lref\vafawit{C Vafa and E. Witten, ``Dual String Pairs with N=1 and
N=2 Supersymmetry in four dimensions,'' hep-th/9507050.}
\lref\gao{H. Gao, ``More Dual String Pairs from Orbifolding,''
hep-th/9512060.}
\lref\chain{G. Aldazabal, A. Font, L. Ibanez, and F. Quevedo,
``Chains of N=2 D=4 heterotic/Type II duals,'' hep-th/9510093.}
\lref\holl{E. Corrigan and T. Hollowood, ``Comments on the Algebra
of Straight, Twisted and Intertwining Vertex Operators,''
{\it Nucl. Phys.}{\bf B304} (1988) 77.}
\lref\duff{M. Duff, ``Strong/Weak Coupling Duality from the Dual String,''
hep-th/9501030, \NP {\bf B442} (1995) 47.}
\lref\mukai{S. Mukai, ``Finite Groups of Automorphisms of K3-Surfaces
and the Mathieu group,'' {\it Invent. math.} {\bf 94} (1988) 183.}
\lref\beauty{L. Dixon, P. Ginsparg, and J. Harvey, ``Beauty and the Beast:
Superconformal Symmetry in a Monster Module,'' {\it Comm. Math.
Phys.} {\bf 119} (1988) 285.}
\lref\flm{I. Frenkel, J. Lepowsky and A. Meurman, ``A natural
representation of the Fischer-Griess monster with the modular function
J as character,'' {\it Proc. Natl. Acad. Sci. USA} {\bf 81} (1984) 3256;
``A moonshine module for the monster,'' J. Lepowsky et al. (eds.),
Vertex Operators in mathematics and physics, Springer Verlag, New York
1985.}
\lref\tuitea{M.P Tuite, ``On the relationship between monstrous
moonshine and the uniqueness of the monster module,''
\CMP {\bf 166} (1995) 495.}
\lref\tuiteb{M.P Tuite, ``Monstrous Moonshine from Orbifolds,''
\CMP {\bf 146} (1992) 277.}
\lref\tuitec{M.P Tuite, ``Generalized Moonshine and Abelian Orbifold
Constructions,'' hep-th/9412036.}
\lref\dongmas{C. Dong and G. Mason, ``On the construction of the moonshine
module as a $\IZ_p$ orbifold,'' {\it Contemporary Math.} {\bf 175} (1994) 37.}
\lref\dongmasb{C. Dong and G. Mason, ``An Orbifold Theory of Genus Zero
Associated to the Sporadic Group $M_{24}$,'' \CMP {\bf 164} (1994) 87;
``Nonabelian Orbifolds and the Boson-Fermion Correspondence,'' \CMP
{\bf 163} (1994) 523.}
\lref\sche{A. N. Schellekens, ``Meromorphic c=24 conformal field theories,''
CERN preprint CERN-TH.6478/92, 1992.}
\lref\dolan{L. Dolan, P. Goddard, and P. Montague, {\it Nucl. Phys.}
{\bf 338} (1990) 529.}
\lref\mont{P.S. Montague, ``Codes, lattices and conformal field theories,''
Cambridge University Ph.D. dissertation, 1991; ``Orbifold constructions
and the classification of self-dual c=24 conformal field theories,''
\NP {\bf B428} (1994) 233.}
\lref\borch{R.E. Borcherds, ``Monstrous moonshine and monstrous Lie
superalgebras,'' {\it Invent. Math.} {\bf 109} (1992) 405.}
\lref\dvvv{R. Dijkgraaf, C. Vafa, E. Verlinde and H. Verlinde, ``The
operator algebra of orbifold models,'' \CMP {\bf 123} (1989) 485.}
\lref\conwaynort{J.H. Conway and S.P. Norton, ``Monstrous Moonshine,''
{\it Bull. London Math. Soc.} {\bf 11} (1979) 308. See also
G. Mason, ``Finite Groups and Modular Functions,'' {\it Proc. Symp. Pure Math.}
{\bf 47} (1987) 181, with appendix by S.P. Norton.}
\lref\conwayex{J.H. Conway, ``Three Lectures on Exceptional Groups,''
in Sphere Packings, Lattices and Groups, J.H. Conway and N.J.A. Norton,
Springer Verlag, New York, 1988.}
\lref\sorba{P. Sorba and B. Torresani, ``Twisted Vertex Operator and String
Theory,'' \IJMP {\bf 3} (1988) 1451.}
\lref\kondo{T. Kondo, ``The automorphism group of Leech lattice and
elliptic modular functions,'' {\it J. Math. Soc. Japan} {\bf 37} (1985) 337.}
\lref\vafa{M. Bershadsky, V. Sadov, and C. Vafa, ``D Strings on D Manifolds,''
hep-th/9510225; ``D-Branes and Topological Field Theories,'' hep-th/9511222;
C. Vafa, ``Gas of D Branes and Hagedorn Density of BPS States,''
hep-th/9511088; H. Ooguri and C. Vafa, ``Two-Dimensional Black Hole and
Singularities of CY Manifolds,''hep-th/9511164.}
\lref\lian{B. Lian and S.-T. Yau, ``Mirror Maps, Modular Relations and
Hypergeometric Series 1,2,'' hep-th/9507151, hep-th/9507153;
``Arithmetic Properties of the mirror map and quantum coupling,''
hep-th/9411234.}
%
%Title page
%
\baselineskip 12pt
\Title{\vbox{\baselineskip12pt
\hbox{NSF-ITP-95-152}\hbox{UCSBTH-95-35} \hbox{\tt hep-th/9512226} }}
{\vbox{\hbox{\centerline{\bf MONSTROUS STRING-STRING DUALITY}}}}
\centerline{\tiau Shyamoli Chaudhuri\footnote{$^\dagger$}{sc@itp.ucsb.edu}}
\vskip .1in
\centerline{\it Institute for Theoretical Physics}
\centerline{\it University of California}
\centerline{\it Santa Barbara, CA 93106-4030}
\vskip .1in
\centerline{\tiau and}
\vskip .1in
\centerline{\tiau David A. Lowe\footnote{$^*$}{lowe@tpau.physics.ucsb.edu}}
\vskip.1in
\centerline{\it Department of Physics}
\centerline{\it University of California}
\centerline{\it Santa Barbara, CA 93106-9530}
\vskip .5cm
\noindent
We analyze the general class of supersymmetry preserving orbifolds of
strong/weak Type IIA/heterotic dual pairs in six dimensions and below.
A unified treatment is given by considering compactification to two
spacetime dimensions and constructing orbifolds by subgroups of the
Fischer-Greiss monster, utilizing the moonshine results of Conway and
Norton. Duality requires nontrivial Ramond-Ramond fluxes on the Type IIA
side which are localized at the fixed points. Further orbifolding by
$(-1)^{F_L}$ gives examples of new four dimensional N=2 Type IIA
vacua which are not conformal field theory backgrounds.

\Date{December 1995}
%\draftmode

\newsec{Introduction}

Of the many duality relationships linking compactifications of the
mysterious eleven dimensional {\bf M} theory perhaps the best
understood are the conjectured string-string dualities, in which a
strongly coupled string theory describes another weakly
coupled string theory \refs{\witt \hullt \duff \hs \sens \kv \fhsv
\joed \petr  \small{--} \jhs}. The evidence for a strong/weak
duality relating the heterotic string compactified on $T^4$ and
the Type IIA string compactified on $K3$ includes the matching of
their moduli spaces \refs{\aspinpt\seiberg{--}\aspmor}, a transformation of
their respective low-energy effective field theory limits
\refs{\witt,\hullt}, and an explicit construction of a nonsingular
heterotic string soliton in the Type IIA theory \refs{\hs,\sens}.
The necessary nonabelian gauge bosons appear as collective
coordinates of the soliton in the Ramond-Ramond sector.

In this paper we provide a general analysis of supersymmetry
preserving orbifolds of strong/weak Type IIA/heterotic dual
pairs in six dimensions and below. The heterotic soliton construction
of Harvey and Strominger, and of Sen, provides a map between target
space duality symmetries of the IIA theory and corresponding
transformations on the world-sheet fields of the heterotic
compactification. Provided such a symmetry is {\it freely acting},
one expects the theories obtained by orbifolding to provide additional
strong/weak dual pairs
\refs{\fhsv,\vafawit,\schsenb,\hls,\clow,\aspin,\aspe{--}\gao}. Thus,
from this duality alone one can derive a remarkable web of strong/weak
dualities in dimensions six and below.

In previous work \clow\ we obtained dual pairs by orbifolding the
IIA theory on $K3$ by an abelian symplectic automorphism \nikulin, i.e., an
automorphism which leaves fixed the holomorphic two-form on $K3$.
Such automorphisms always have {\it fixed points} on $K3$. To ensure that
the symmetry is freely acting, the automorphism is
accompanied by a shift corresponding to a Wilson line for the ten-dimensional
Ramond-Ramond gauge field of the Type IIA theory \refs{\schsenb}.
The duality map determines a corresponding automorphism plus shift of the
Narain lattice on the heterotic side. It is interesting that abelian
symplectic automorphisms of the classical geometry automatically satisfy
level matching \refs{\clow,\aspin}, so that the accompanying shift can
be chosen to be geometrical. This will not be true more generally.

The symplectic automorphisms of the classical geometry of
$K3$, including nonabelian cases, have been classified by
Mukai \refs{\mukai}. He notes that they are in one-to-one
correspondence with the purely
left-moving symmetries of the $(19,3)$ Lorentzian self-dual
lattice formed by $H^2(K3,\IZ)$, fitting into special subgroups of
the finite sporadic group $M_{23}$. An obvious extension is to consider
automorphisms of the quantum geometry of the IIA theory on $K3$, and
its further toroidal compactifications.

A new feature of the orbifolds we consider is the possibility of
Wilson line backgrounds for the 24 Ramond-Ramond $U(1)$ gauge fields
of the IIA theory on $K3$. Gauge transformations in these
$U(1)$ gauge fields are mapped under duality to general shifts
in the $\Gamma^{(20,4)}$ Narain lattice of the heterotic string on
$T^4$. By considering orbifolds with respect to these more
general symmetries, combined with the action of a symplectic
automorphism, we obtain Type IIA theories with RR fluxes localized
at the fixed points of $K3$. We find that there is considerable
freedom in the choice of RR fluxes. Each choice is mapped under
duality to a specific shift vector on the heterotic side. The
different possibilities for the shift vector give otherwise
identical low energy limits, and are therefore associated with
T-duality symmetries. It should be noted that from the viewpoint
of the Type IIA compactification, these are T-duality symmetries
of the Ramond-Ramond sector and do not originate in conformal
field theory!

We will show how all these orbifolds
fit into a unified structure by considering compactifications
down to two dimensions. We make the natural conjecture that
the supersymmetry preserving symmetries of the Type II theory on
$K3\times T^4$
correspond to the purely left-moving symmetries of a $(24,8)$ self-dual
Lorentzian lattice.
We make an additional restriction that the lattice takes the
form $\Gamma^{(24,0)} \oplus \Gamma^{(0,8)}$. Such a point in the
moduli space can always be reached on the Type II side by turning
on suitable $B$ field deformations. The
24-dimensional Euclidean lattices have been classified
and are the Leech lattice and its cousins the 23 other Niemeier lattices
\niem.
The other Niemeier lattices lead to heterotic theories with enhanced
gauge symmetry, which are expected to be dual to Type II theories
on singular $K3 \times T^4$ manifolds. To avoid such subtleties we
will focus mostly on orbifolds of the Leech lattice.

The automorphism group of the Leech lattice
is one of the simple finite groups $\cdot 0$, the
Conway group. This group is closely related to the monster group $\IM$,
the largest of the sporadic groups. The group $\IM$ is
the automorphism group of the monster module
conformal field theory $V^\natural$ obtained
as a $\IZ_2$ orbifold of the Leech lattice theory, where
the $\IZ_2$ acts by changing the sign of the coordinates on the lattice.
This $\IZ_2$ orbifold of the
Leech lattice theory contains no massless states in its twisted sectors --
therefore we expect there will be a dual IIA theory, obtained via
a freely acting orbifold of the quantum geometry. One may then
consider further orbifolds of the theories based on $V^\natural$,
or simply the Leech lattice, yielding
a large class of candidate dual pairs with maximal supersymmetry.
Note that, in contrast to the approach of \refs{\kv,\chain},
the soliton string construction gives an unambiguous method to
obtain reliable dual pairs as orbifolds by freely acting
symmetries.

Having established a duality map in two spacetime dimensions we
can lift the action of the supersymmetry preserving automorphism
into higher dimensions by decompactifying any invariant
toroidal coordinates. In particular, all of the six dimensional
classical orbifolds of \refs{\schsenb,\clow} can be recovered by
this procedure.\foot{We are assuming that the $B$ field may be turned
off in a way that preserves the symmetry. To prove this statement
requires a more detailed knowledge of the moduli space of quantum
$K3$ surfaces than is presently available.} As mentioned above,
all of the classical $K3$ automorphisms fit into subgroups of $M_{23}$,
leaving invariant an $\Gamma^{(1,1)}$ component of the $(20,4)$
lattice. This is not true for the quantum automorphisms--- we find
symplectic automorphisms that lie in $M_{24}$, but not in $M_{23}$.
We also find quantum symplectic symmetries of $K3$$\times$$T^4$ that require
a further extension to the Conway group.

The plan of the paper is as follows. In section 2 we clarify
the duality dictionary between symmetries of the toroidal
left-moving conformal field theory on the heterotic side and
symplectic symmetries of $K3$ on the Type II side. In section 3
we consider dual Type IIA/heterotic compactifications
to two dimensions. Using the monstrous moonshine results of
Conway and Norton, we present the general construction for
cyclic orbifolds of the Leech lattice and the Moonshine module,
giving an implicit definition of Type IIA/heterotic duals which
cannot be obtained by a consideration of classical symmetries
of $K3$ alone. In Section 4, we explain how our construction can
be easily modified to obtain $N$$=$$2$ Type IIA/heterotic dual
pairs. We conclude with some comments and a brief discussion
of our results.

\newsec{The String-String Duality Dictionary}

We begin by considering the strong/weak coupling duality between the
heterotic string on $T^4$ and the Type IIA string on $K3$.
The heterotic compactification is described by a Narain lattice
$\Lambda=\Gamma^{(20,4)}$, i.e. an even self-dual Lorentzian lattice.
Using the construction of the heterotic string as a soliton
of the IIA theory \refs{\hs,\sens}, duality maps this lattice to a point in the
moduli space of the IIA string on $K3$.
Likewise, symmetries of the heterotic CFT will be mapped to
symmetries of the IIA theory. To see this explicitly, we need
the expressions for the zero modes of the soliton, which give
rise to the worldsheet fields of the heterotic string. Zero modes
coming from the Ramond-Ramond (RR) three-form potential are
\eqn\zmodes{
C= {\alpha' \over{2\pi}} X^I(\sigma) U_I(y) \wedge de^{2\phi_0-2\phi(x)}~,
}
in the notation of \hs, with $\phi(x)$ the background dilaton field.
The $U_I(y)$ are the harmonic two-forms of $K3$, with $I=1,\cdots,22$.
The worldsheet fields satisfy the chiral constraint
\eqn\cconst{
*U_I = U_J H^J_I~,
}
where $H$ has signature $(19,3)$, yielding 19 left-moving and 3 right-moving
worldsheet coordinates. An additional zero mode comes from
\eqn\zzmode{
A={1\over {2\pi}} X^0(\sigma) d e^{2\phi_0-2\phi(x)}, \quad
C=-{1\over \pi}  X^0(\sigma) e^{2\phi_0-2\phi(x)} H~,
}
which gives a single (left,right)-moving bosonic zero-mode. Here
$A$ is the RR one-form potential of the IIA theory in ten dimensions,
and $H$ is the three-form field strength. There are further
bosonic zero modes arising from transverse motion in six-dimensional
Minkowski space, together with fermionic zero modes, but these will not
be relevant in the following discussion.

Let us see how the duality dictionary
works for the Type II dual of the simplest
of the CHL models \refs{\chl,\schsenb}. On the heterotic side, the model
is constructed as an orbifold by a $\IZ_2$ symmetry which
exchanges the two $E_8$ components of the lattice and acts
with a half period shift on one of the circles of the torus
$T^4$. On the Type II side, the exchange of the $E_8$'s corresponds
to a certain symplectic automorphism of $K3$. \foot{Note that
it is possible to move away from this point of enhanced gauge symmetry,
preserving the $\IZ_2$ symmetry, so that $E_8 \times E_8$ is broken
down to $U(1)^{16}$. Thus one need not consider singular $K3$ surfaces.}
The shift is identified with a $\IZ_2$ gauge transformation of the
$U(1)$ Ramond-Ramond (RR) gauge field $A$.
The resulting orbifold on the Type II side
is not a conventional superconformal field theory. A nontrivial
Wilson line background of the RR gauge field is present, with fluxes
localized at the fixed points of the $K3$.

There are a total of 24 RR $U(1)$ gauge fields present in the IIA
theory on $K3$. One comes from the 10d RR  $U(1)$ gauge field as
mentioned above, one arises from the dual of the RR four-form field strength,
and 22 arise from the RR three-form potential integrated over
a nontrivial two-cycle of $K3$, i.e. we may write
\eqn\rrgauge{
C= U_I \wedge A^I~.
}
It is natural therefore to consider
orbifolds which turn on Wilson line backgrounds in these
other RR gauge fields.
It is clear from \zmodes,\zzmode\ that these nontrivial
gauge transformations are equivalent to shifts in the $X^I$ and $X^0$.
On the heterotic side, the inclusion of these Wilson lines
will correspond to orbifolds involving general shifts
of the $\Gamma^{(20,4)}$ lattice.

In the following section, we will combine these general shifts
with supersymmetry preserving lattice automorphisms to
construct a large class of maximally supersymmetric dual pairs.
The classical symplectic automorphisms of $K3$ will map to
lattice automorphisms \mukai. However the complete set of lattice
automorphisms are expected to map onto more general quantum
symmetries of the IIA string on $K3$ which are difficult to
describe explicitly. However, from the heterotic point of view,
these quantum symmetries are precisely on the same footing
as the classical symmetries. This suggests that the Type II orbifolds
by these quantum symmetries will be implicitly defined by their
heterotic counterparts.

So far, we have discussed the mapping of symmetries under duality
from the heterotic string on $T^4$ to the IIA string on $K3$.
Clearly the same kind of arguments will go through when one
considers further toroidal compactification on an additional
$T^4$, as described in the following section.

\newsec{Supersymmetry Preserving Orbifold Dual Pairs}

We will begin by considering compactifications to two spacetime
dimensions described by $(24,8)$ dimensional Lorentzian self-dual
lattices\foot{This construction was suggested to us
by J. Harvey.}
\eqn\latone{
\Gamma^{(24,8)} = \Lambda \oplus \Gamma^8~,
}
where $\Gamma^8$ is the $E_8$ lattice. For the left moving lattice, we
take the rank $24$ self-dual lattice $\Lambda_{A_1^{24}}$.
This lattice consists of the $(SU(2))^{24}$ root lattice plus
the conjugacy classes (doublets)
obtained from the following generators \niem,
\eqn\weights{
( 1, [ 0,0,0,0,0, 1,0, 1,0,0, 1, 1, 0,0,1, 1,0, 1,0,1,1,1, 1 ])~,
}
where the square brackets denote cyclic permutations.
The Leech lattice, $\Lambda_{Leech} $, is easily generated from
$\Lambda_{A_1^{24}}$ as an orbifold by the purely left moving $\IZ_2$ shift
\eqn\tran{
h : X \to X + 2 \pi \cdot {1\over 4} \sum_{i=1}^{24} a_i~,
}
where the $a_i$ are the $24$ positive roots of $(SU(2))^{24}$
\refs{\beauty}. Since the Leech lattice has no vectors of length squared
two the only massless states in this conformal field theory are the
$24$ $U(1)$ bosonic states $| \alpha^i_{-1} >$.

The automorphism group of this conformal field theory is known as
the Conway group, which is a close relative of the monster
group $\IM$, the largest of the sporadic groups. The group $\IM$ is
the automorphism group of the monster module conformal field theory,
obtained by the $\IZ_2$ involution
\eqn\invol{
g: X \to -X~,
}
for all $24$ coordinates on the Leech lattice. The monster module
conformal field theory has no massless states in its twisted
sectors. Thus, for each supersymmetry preserving automorphism of the
monster module we expect a dual IIA theory obtained via a freely
acting orbifold of the quantum geometry.

We will consider supersymmetry preserving automorphisms of the
two, closely related, conformal field theories based on these
lattices. Many of our results extend to more general lattices with
signature $(24,8)$. Our restriction is helpful for technical reasons.
We will be interested in symmetries which act purely on the
left-moving coordinates of the lattice. Orbifolds will be
constructed by combining the action of these symmetries with
shifts with right-moving components, ensuring no massless states
appear in the twisted sectors. For a subclass of these theories
the Type II dual may be explicitly constructed \clow\ as a
{\it geometric} orbifold of the IIA string on $K3$ with a RR Wilson
line turned on, but more generally the IIA theory is a nongeometric
orbifold of a $K3$ compactification, as discussed above.

\subsec{Cyclic Orbifolds of the Leech Lattice CFT}

Begin by considering the group of automorphisms of the Leech
lattice conformal field theory $\hat \Lambda$.
Cyclic orbifolds of the holomorphic CFT based on the Leech lattice
have previously been considered in
\refs{\tuitea\tuiteb\tuitec\mont{--}\dongmas}.
Closely related constructions based on other $c=24$ holomorphic CFTs
have been considered in \refs{\dongmasb,\sche,\dolan}.
We denote the Leech lattice by $\Lambda$. As explained more
generally above, the automorphism group of the Leech lattice
conformal field theory contains the Conway group $\cdot 0$, the
automorphism group of the Leech lattice \conwayex, but is extended by the
automorphism group of the cocycle operators.

A highest weight state in the untwisted Hilbert space is constructed as
\eqn\leechst{
|\beta \rangle = V(\beta,0) |0\rangle~,
}
where $\beta\in \Lambda$ and the vertex operator $V(\beta,z)$ is
defined as
\eqn\vertop{ V(\beta,z) = : e^{i\beta \cdot x(z)} : c(\beta)~,
}
where $c(\beta)$ is an element in the central extension of
$\Lambda$ by $\IZ_2$. Associativity of the vertex operator algebra
requires that the $c(\beta)$ satisfy cocycle conditions
\eqn\cocyc{
\eqalign{
c(\alpha) c(\beta) &= \epsilon(\alpha, \beta) c(\alpha+\beta) \cr
\epsilon(\alpha,\beta) &= (-1)^{\alpha\cdot \beta}
\epsilon(\beta,\alpha) ~.\cr
}}

For $\bar a \in \cdot 0$, $\bar a: \Lambda \to \Lambda$ and the
metric is preserved $\bar a \alpha \cdot \bar a \beta = \alpha
\cdot \beta$. The action on the untwisted  Hilbert space is simply
$|\beta\rangle \to | \bar a \beta \rangle$. However, due to the
cocycle factors that appear in \vertop\ the full automorphism
group of $\hat \Lambda$ is the central extension of $\cdot 0$
by $\IZ_2^{24}$ which is denoted $C_0= 2^{24} (\cdot 0)$.
For $a\in C_0$ the action on the CFT is
\eqn\coact{
\eqalign{
a: c(\beta) &\to (-1)^{f_a(\beta)} c(\bar a \beta) \cr
a: |\beta \rangle &\to (-1)^{f_a(\beta)} | \bar a \beta \rangle~, \cr}
}
where $\bar a$ is the element in $\cdot 0$ corresponding to $a$,
and $f_a(\beta)$ is a $\IZ_2$ central extension of $\bar a$, with
$f_a(\alpha+\beta) = f_a(\alpha) + f_a(\beta)$. A useful
representation is
\eqn\frep{
f_a(\beta)= \beta \cdot m~,
}
where $m$ lies on the dual lattice. If $e_r$ is a basis for
the dual lattice, then $m= m^r e_r$ with $m^r = 0,1$.

Associated to any element $\bar a$ of $\cdot 0$ of order $n$
is a characteristic polynomial
\eqn\cpol{
{\rm det} (x-\bar a) = \prod_{k|n} (x^k-1)^{a_k}~,
}
where
\eqn\ksum{
\sum_k  k a_k =24~,
}
and the $a_k$ are integers. The
symbol $\prod_{k|n} k^{a_k}$ is known as the Frame shape of
the element.  Frame shapes for all elements in $\cdot 0$ are tabulated in
\kondo. If $\bar a$ lies in the $M_{24}$ subgroup of $\cdot 0$,
which has a representation as permutations of 24 letters,
then the Frame shape is simply the cycle decomposition of the
permutation.

Now we are ready to consider $\IZ_n$ orbifolds by elements
in $C_0$. We first consider the case when $m=0$.
In general, when a symmetry $a\in C_0$ is lifted to a symmetry
of the orbifold the order of $a$ may increase to $2n$ when
acting on states in the twisted sector. This point is discussed
for example in \sorba. The result is that $a$ will be of order
$n$ provided $n \beta_{a}^2 = 0~{\rm mod}~2$, where $\beta_a$ is
the component of $\beta$ invariant under $a$
\eqn\invec{
\beta_a = 1/n \sum_{k=0}^{n-1} a^k \beta~.
}
This condition holds for all the automorphisms of the Leech
lattice \tuiteb\ thus provided $m=0$, the action of $a$ in the
twisted sector will always be of order $n$.

The vacuum energy of the left-movers is
\eqn\lvac{
E_L= -1 + {1\over 4} \sum_{i=1}^{24} r_i (1-r_i)~,
}
where $\exp(2\pi i r_i)$ are the eigenvalues of coordinates of
$\Lambda$ under the action of $\bar a$ in a diagonal basis. Using
\ksum\ we find
\eqn\lvacis{
E_L = - {1\over 24} \sum_{k|n} {a_k \over k}~.
}
To obtain a perturbatively consistent theory we need to impose
the level-matching condition
\eqn\levmat{
n(- {1\over 24} \sum_{k|n} {a_k \over k} + {1\over 2} \delta^2)
= 0~{\rm mod}~1~,
}
where $\delta^2 = \delta_L^2-\delta_R^2$.

It appears it is possible to construct a shift vector of order
$n$ which satisfies this condition for every element in $\cdot 0$.
A sample of such orbifolds is listed in the table in appendix A.
The first six entries recover all of the cyclic abelian automorphisms
appearing in Nikulin's classification of symplectic symmetries of
classical $K3$ surfaces, and were obtained previously as orbifold
dual pairs in six and four dimensions \refs{\schsenb,\clow,\aspin}.
They satisfy level matching without additional shift vectors, though
here we include a shift to eliminate massless states in the
twisted sectors.
The next four entries are quantum $K3$ automorphisms.
All of the above automorphisms, and the
extension to cyclic products of these automorphisms $\prod_{k} \IZ_{N_k} $
considered by Mukai fit within the finite group $M_{23}$ .

The next seven entries of the table are quantum $K3$
symmetries that fit within the finite group $M_{24}$,
but not $M_{23}$. A difference with the symmetries already considered
is that now a nontrivial shift vector is required to
satisfy level matching. All of these examples contain radial moduli
which allow us to decompactify smoothly to six dimensional theories.
The remaining entries-- including the $\IZ_{20}$ and $\IZ_{60}$ elements
contained in the Conway group, but not in $M_{24}$-- are examples of Leech
lattice orbifolds that cannot be decompactifed to six dimensions.
Note that there is considerable freedom in the choice of the shift
vector, giving identical low-energy theories. Under the duality map,
this corresponds to the freedom in choosing Ramond-Ramond fluxes
in the Type IIA theory. This gives rise to T-duality symmetries of
the Type IIA theory which do not originate in conformal field theory!

More generally, one may consider an element $a$ in $C_0$
corresponding to nonzero $m$. Now one finds that the action
of $a$ on states in the twisted sectors is of order $2n$ in general
\tuiteb. The vacuum energy of the left-movers is modified to
\eqn\nlmov{
E_L = - {1\over 24} \sum_{k|n} {a_k \over k} + {1 \over 8} m_a^2~.
}
The level-matching condition now becomes
\eqn\levmatn{
2n(- {1\over 24} \sum_{k|n} {a_k \over k} + {1 \over 8} m_a^2 +
{1\over 2} \delta^2 ) = 0~{\rm mod}~1~.
}
A simple example of such an orbifold is to take $\bar a= 2^{12}$ and
$m=(1,0^{23})$. This is to be combined with a right-moving order
4 shift $\delta= (2,1,1,1,1)/4$.

\subsec{Monstrous moonshine}

Before proceeding to orbifolds of the monster module CFT, let us
review the monstrous moonshine conjectures of Conway
and Norton \conwaynort, which have subsequently been proven by
Borcherds \borch, and establish some notation and terminology
which will be useful later. A helpful review of many of these ideas
aimed at string theorists may be found \beauty.
Consider the fundamental domain of the modular group
$\IH/ SL(2,\IZ)$, where $\IH$ is the upper half plane.
Compactifying this space by adjoining the point at infinity,
one obtains a space with the topology and complex structure of the
Riemann sphere. The modular function $j$ gives a one-to-one and onto
map from $\IH/ SL(2,\IZ) \cup i\infty$ to the Riemann sphere
$\IC \cup \infty$,
thus the Riemann surface $\IH/ SL(2,\IZ) \cup i\infty$ has genus zero.
This means that the field of meromorphic functions on
$\IH/ SL(2,\IZ) \cup i\infty$ is just the rational functions of $j$
with complex coefficients. $j$ is said to be the {\it hauptmodul}
of this function field. It is also true that other discrete
subgroups $\Gamma$ of $SL(2,\IR)$ yield
genus zero Riemann surfaces as the compactification of
$\IH/ \Gamma$. Associated with these subgroups are hauptmoduls
$j_\Gamma$ which generate the genus zero function field.

A Thompson series for an element $g$ of the monster $\IM$ is defined
as
\eqn\thomp{
T_g(q) = q^{-1} {\rm Tr}g q^{L_0}~,
}
where the trace runs over all the states in the monster module
$V^\natural$. As discussed in \flm, $V^\natural$ may be constructed
as a $\IZ_2$ orbifold of the Leech lattice conformal field theory, where
the $\IZ_2$ acts by changing the sign of all the left-moving coordinates.
Conway and Norton conjectured that the $T_g$ are hauptmoduls for some
genus zero subgroup $\Gamma_g$ of $SL(2,\IR)$. Properties of these
subgroups and hauptmoduls may be found for each conjugacy
class of the monster in the tables of \conwaynort. For
$g$ of order $n$, $T_g$ is fixed by the subgroup $\Gamma_0(n)$
up to $h^{\rm th}$ roots of unity, where $h|n$ and $h|24$.
This subgroup is defined by
\eqn\consub{
\Gamma_0(n) = \bigl\{ \bigl( \matrix{ a & b \cr c & d \cr} \bigr)
\in SL(2,\IZ) \big| c=0~{\rm mod}~n \bigr\}~.
}
$T_g$ is invariant under $\Gamma_0(N)$ with $N=nh$. The maximal
fixing group $\Gamma_g$ lies between $\Gamma_0(N)$ and
the normalizer of $\Gamma_0(N)$ in $SL(2,\IR)$. The normalizer
of a subgroup $H$ of $G$ consists of the elements $g\in G$
such that $g^{-1} h g \in H$ for all $h\in H$.
This normalizer contains the Fricke involution
($w_N:\tau \to -1/N \tau$). The conjugacy classes
with $h=1$ will be referred to as {\it normal}, the
others as {\it anomalous}. The maximal subgroup of $SL(2,\IR)$
which fixes $T_g$ up to an $h^{\rm th}$ root of unity
will be called the eigenvalue group $E(g)$.
Also we will refer to classes
which contain the Fricke involution in $E(g)$ as {\it Fricke}
classes.

In the notation of \conwaynort, $E(g)$ is normal
when it is of the form $n+e,f,g,\cdots$. The $n$ denotes
the group $\Gamma_0(n)$ and the $e,\cdots$
denote Atkin-Lehner involutions $w_e$ adjoined onto $\Gamma_0(n)$.
$E(g)$ is anomalous when it is of the form $n|h +e,f, \cdots$.
The symbol $n|h$ denotes the group of matrices $\Gamma_0(n|h)$
\eqn\nhmat{
\bigl( \matrix{ a & b/h \cr cn &d} \bigr)~,
}
with determinant one. For $\Gamma_0(n|h)$ the Fricke
involution is $w_{n/h}$. The $w_e$ (where $e\in \IZ$) for
$\Gamma_0(n|h)$ are defined as the set of matrices
\eqn\wedef{
\bigl( \matrix{ ae & b/h \cr cn &de} \bigr)~,
}
with determinant $e$, $a,b,c,d \in \IZ$ and $e || n/h$ (i.e.
$e$ divides $n/h$ and the greatest common divisor $(e, n/he) =1$).

\subsec{Cyclic orbifolds of the Monster Module CFT}

Now let us consider cyclic orbifolds of the monster module conformal
field theory.
All the orbifolds we obtain this way will have the maximal
rank reduction 24, and will only exist in two spacetime dimensions.
The simplest type of elements
to consider are the non-Fricke normal elements. There
are 38 such conjugacy classes in the monster. The prime
order $p$ classes of this type satisfy $p-1|24$, and
have $\Gamma_g = \Gamma_0(p)$. It turns out
for each of these 38 classes there is a corresponding element
in $\cdot 0$, so we may think of these orbifolds as
simply automorphisms acting on the Leech lattice, which
may then be lifted to $V^\natural$.
The partition function in the sector twisted by $g$
is obtained by acting with $S: \tau \to -1/\tau$ on $T_{g^{-1}}$,
\eqn\tpart{
Z(1,g) = q^{-1} {\rm Tr}_{H_g} q^{L_0} = T_{g^{-1}}( S q)~.
}
Since the element is normal, $Z(1,g)$ is invariant under
$\tau \to \tau+n$, therefore the twisted sector satisfies level
matching.
We wish to accompany the action of $g$ by a shift, to ensure that
no massless states appear in the twisted sectors. For each
of the 38 classes it is possible to construct a purely right-moving
shift that acts in the $\Gamma^{(0,8)}$ component of the lattice,
and which satisfies level matching $n \delta^2/2 = 0~{\rm mod }~1$.

Now let us consider orbifolds by the 82 normal Fricke classes of
the monster. In general, these symmetries do not lie in the $\cdot 0$
subgroup of the monster. However it is still true that the
partition function in the sector twisted by $g$ is given
by a Thompson series as in \tpart, so we will assume the
twisted sector ${\cal H}_g$ is well-defined. The general
construction of orbifolds based on such nongeometrical symmetries
has been studied in detail in \dvvv.
Since the element
is normal, the action of $g$ satisfies level matching. The shifts
are constructed as in the preceding paragraph.

Orbifolds generated by the 51 anomalous classes of the monster are
somewhat more subtle. The twisted sector does not satisfy level
matching unless the action of the symmetry is combined with a
nontrivial purely right-moving shift vector $\delta$.
Consider an element $g \in \IM$ with eigengroup $n|h + e,f,\cdots$.
Under the transformation $\tau \to \tau+n$, the
partition function for the twisted sector without a shift
picks up a phase $e^{2\pi i l/h}$, where $l$ is some integer.
The global phase anomaly is canceled by combining the action
of $g$ with a shift $\delta$ which satisfies
$n( \delta^2/2 + l/h) = 0~{\rm mod}~1$. Such shifts may be
constructed for each element of the monster, however the analysis
is performed case by case.

We present some examples of these
orbifolds in the table in appendix B.

\newsec{New N=2 Type IIA/Heterotic Dual Pairs}

It is straightforward to obtain
dual pairs with $N=2$ supersymmetry in 4d from the subclass of maximally
supersymmetric dual pairs which may be decompactified to six dimensions.
Within this class,  before orbifolding to obtain the reduced $N=4$ theory,
there is a region in moduli space on the heterotic side in which the
Narain lattice decomposes as
\eqn\hlat{
\Gamma^{(24,8)} = \Gamma^{(20,4)} \oplus \Gamma^{(4,4)}~.
}
On the Type II side, where we start with a compactification on
$K3\times T^2$, we may orbifold by the $(-1)^{F_L}$
symmetry combined
with a $\IZ_2$ shift in the $T^2$ satisfying $\delta^2=0$,
along the lines of the constructions in \refs{\hls,\vafawit}.
As discussed in these references, $(-1)^{F_L}$ acts as a change
in sign of the $\Gamma^{(20,4)}$ component of the lattice on the
heterotic side. To satisfy level matching an additional shift in
the $\Gamma^{(4,4)}$ component of the lattice is required,
satisfying $\tilde \delta^2 = 1/2$.
On each side, these $\IZ_2$ orbifolds break the supersymmetry
down to $N=2$. One may then proceed with the construction
of reduced theories as described in the previous sections,
orbifolding by some discrete group $G$. In general,
one typically needs to also check level matching is satisfied
in the sectors twisted by the product of $\IZ_2$ with an element in $G$.
In these theories one is left with a low energy spectrum which
contains four $U(1)$ gauge fields at a generic point and
$20- \Delta r$ hypermultiplets, where $\Delta r$ is the gauge group
rank reduction of the corresponding $N=4$ theory.

The resulting Type II theories will have $(1,4)$ worldsheet
supersymmetry. As discussed in \vafawit\ the dilaton in such
theories lies in a vector multiplet. The dilaton on the
heterotic side likewise lies in a vector multiplet. In each case
we therefore expect quantum corrections to the vector multiplet
moduli space. This class of dual pairs does not display
second-quantized mirror symmetry \fhsv. The hypermultiplet
moduli space does not receive quantum corrections on either
side. Unlike the examples considered in \kv , it is possible to
obtain hypermultiplet moduli spaces of low dimension. We
plan to return to these models in future work.

We emphasize that the four dimensional $N$$=$$2$ Type IIA vacua
we are considering do {\it not} have a straightforward conformal
field theory description since they involve Ramond-Ramond
backgrounds. Thus, they are also the simplest examples of the
new Type IIA vacua proposed in \refs{\pols}. The backgrounds
are described by electromagnetic flux configurations, localized
at fixed points on the compactification.

\newsec{Conclusions}

Our analysis has been restricted to the simplest Type IIA vacua
which lie outside of conformal field theory, and to phenomena that
occur in the generic moduli space of these theories. More generally,
Type II string theories may contain a variety of both soft
\refs{\chs,\jeff,\hors}
and hard \refs{\dlp,\joep} BPS saturated p-brane solitons, including
those with RR charge. In a beautiful recent insight \refs{\joep}, it
has been shown that the quantum degrees of freedom required by string
duality are the Dirichlet-branes of a Type I string theory \refs{\dlp}.
D-branes potentially provide a powerful calculus for the dynamics
of BPS states in string theory. They should provide a much more
explicit description of nonperturbative string effects, such as
nonabelian gauge symmetry, associated with regions of the moduli
space where the Type IIA theory is strongly
coupled \refs{\andyc,\aspinpt,\joed, \vafa,\small}.

An essential ingredient underlying our work was the
resolution in \refs{\schsenb} of the paradox posed by the existence
of a six dimensional $N$$=$$2$ heterotic model with eight fewer
abelian gauge fields at generic points in the moduli space
\refs{\chl,\chpol} than the usual model obtained via toroidal
compactification. The Type IIA dual cannot correspond to a
conformal field theory compactification \refs{\seiberg}. It is
described instead by compactification on a $K3/{\sigma}$ orbifold
with a Ramond-Ramond $\IZ_2$ flux localized at fixed points of
the involution $\sigma : U^i \to - U^i$, $i=1$, $\cdots$, $8$ for
eight of the harmonic forms on the $K3$ surface \refs{\schsenb}. We have
shown the necessity of extending the class of Ramond-Ramond
electromagnetic flux configurations considered on the Type II side
to accommodate the generic strong/weak Type IIA/heterotic orbifold
dual pair in six dimensions and below. We suspect this is only
the tip of the iceberg, and that much remains to be discovered
about generic Ramond-Ramond backgrounds and their relationship to
Type IIA/heterotic duality.

In particular, consider the $19$ anti-self dual (ASD) harmonic forms on
the $K3$ surface. There is a one-to-one correspondence between
the ASD forms, defining the left moving part of the integral
cohomology lattice, and Ramond-Ramond gauge fields which arise from
integrating the RR three-form potential over the associated non-trivial
two-cycles of the $K3$. Consider a compactification $K3/G$, where we
introduce Ramond-Ramond $\IZ_2$ gauge transformations
associated to directions of the lattice for which $G$ has some symplectic
action.  Such an action
is mapped under duality to an automorphism of the heterotic conformal field
theory which introduces {\it nonabelian} discrete torsion, reminiscent
of the fermionic construction of \refs{\chl}. \foot{A similar phenomenon
occurs in the Type I theory \refs{\joed}.
We would like to thank J. Polchinski for pointing out this connection.}

To see the relation to the fermionic models of \chl\ recall that
the automorphism group of a conformal field theory is larger than
that of the underlying lattice. For an $r$ dimensional lattice,
this extension is the $({\IZ}_2)^r$ automorphism group of the cocycles
required by associativity of the operator product algebra \refs{\beauty}.
Automorphisms of the lattice which only satisfy level matching when
accompanied by nontrivial action on the cocycles are associated with
twisted vertex operator algebras \refs{\holl,\sorba}. These are
straightforwardly realized on the world-sheet by Majorana-Weyl
fermions \refs{\chl}. This resolves the puzzle that not all of the
fermionic models of \chl\ were obtained as orbifolds with respect to
abelian lattice automorphisms plus shifts in an invariant direction
\refs{\chpol,\clow}.

In this paper we have explored an algebraic structure contained in
the monster group underlying Type II/heterotic duality.
Recent work has uncovered other connections to the monster.
As discovered by Harvey and Moore \bps , the monster Lie superalgebra
miraculously appears in the algebra of the tower of elementary BPS string
states of a four dimensional $N$$=$$2$ heterotic string theory. They
have further speculated that different dual theories are simply alternate
representations of an underlying algebraic structure, much like the different
realizations of an affine Lie algebra. A related observation has been
made by Lian and Yau who have found that Thompson series are related to $K3$
mirror maps \lian.  Further elucidating these connections will no doubt
lead to a better understanding of strong/weak duality in string theory.

\bigskip
{\bf Acknowledgements}

We thank P. Aspinwall, P. Berglund, J. Polchinski, R. Schimmrigk,
and A. Strominger for useful comments, and C. Dong and G. Mason
for correspondence. This work grew out of discussions with J.
Harvey, to whom we are grateful for numerous helpful suggestions
and references. This work is supported in part by NSF Grants
PHY 91-16964 and PHY 94-07194.
\vskip 1cm

\appendix{A}{Examples of cyclic orbifolds of the Leech lattice CFT}

Column 1 gives the order of the automorphism. The second column
gives the Frame shape of the element in the Conway group.
If the element lies in the $M_{24}$ subgroup of $\cdot 0$,
then the Frame shape is simply the cycle decomposition of the
permutation.
The third column gives an example of a purely right-moving shift
vector required for level matching. All of the examples, except for
the last, contain radial moduli. By decompactifying with respect to
these moduli higher dimensional theories are obtained. The fourth
column gives the maximum number of noncompact dimensions that may be
obtained this way, such that the theory still possesses a
Type IIA dual. Finally, the fifth column gives the overall gauge
multiplet rank reduction.
\vfill
\vbox{
\begintable
$\IZ_n$| $g$| $\delta_R$ | $d_{max}$ | $-\Delta r$ \cr
$\IZ_2$| $1^8 2^8$ | $(\half,\half,\half,\half)$ | 6 | 8 \nr
$\IZ_3$| $1^6 3^6$ | $({2\over 3},{1\over 3}, {1\over 3},0 )$| 6 | 12 \nr
$\IZ_5$| $1^4 5^4$ | $({3\over 5}, {1\over 5},0,0)$ | 6 | 16 \nr
$\IZ_6$| $1^2 2^2 3^2 6^2$ | $({3\over 6}, {1\over 6}, {1\over 6},
  {1\over 6})$ | 6 | 16 \nr
$\IZ_7$| $1^3 7^3$ | $({3\over 7}, {2\over 7}, {1\over 7},0)$ | 6 |18\nr
$\IZ_8$| $1^2\cdot 2 \cdot 4 \cdot 8^2$ | $({3\over 8},{2\over 8},
  {1\over 8},{1\over 8} ,{1\over 8})$ | 5 | 18 \cr
$\IZ_{11}$| $1^2 11^2$ | $({4\over 11}, {2\over 11},{1\over 11},
  {1\over 11})$ | 6 | 20 \nr
$\IZ_{14}$| $1\cdot 2\cdot 7\cdot 14$ | $({5\over 14},{1\over 14},
  {1\over 14},{1\over 14})$ | 6 | 20 \nr
$\IZ_{15}$| $1\cdot 3 \cdot 5 \cdot 15$ | $({5\over 15},{2\over 15},
  {1\over 15},0)$ | 6 | 20 \nr
$\IZ_{23}$| $1\cdot 23$ | $({6\over 23},{3\over 23},{1\over 23},0)$ |
  4 | 22 \cr
$\IZ_{2}$| $2^{12}$ | $({1\over 2},{1\over 2}, 0,0
  )$ | 6 | 12 \nr
$\IZ_{3}$| $3^8$ | $({2\over 3},0,0,0)$ | 6 |
  16 \nr
$\IZ_{4}$| $2^4 4^4$ | $({1\over 4},{1\over 4},{1\over 4},{1\over 4}
  )$ | 6 | 16 \nr
$\IZ_{4}$| $4^6$ | $({2\over 4}, {1\over 4}, {1\over 4}, 0
  )$ | 6 | 18 \nr
$\IZ_{6}$| $6^4$ | $({2\over 6}, {2\over 6}, {1\over 6},{1\over 6}
  )$ | 6 | 20 \nr
$\IZ_{10}$| $2^2 10^2$ | $({3\over 10},{1\over 10},0,0
  )$ | 6 | 20 \nr
$\IZ_{12}$| $2\cdot 4 \cdot 6 \cdot 12$ | $({3\over 12},
  {1\over 12}, {1\over 12}, {1\over 12})$ | 6 | 20 \cr
$\IZ_{4}$| $1^4 2^2 4^4$ | $({2\over 4}, {1\over 4}, {1\over 4},
  {1\over 4},{1\over 4})$ | 5 | 14 \nr
$\IZ_{12}$| $12^2$ | $({4\over 12},{2\over 12},{1\over 12},{1\over 12}
  )$ | 4 | 22 \nr
$\IZ_{21}$| $3\cdot 21$ | $({5\over 21},{1\over 21},{1\over 21},{1\over 21}
  )$ | 4 | 22 \cr
$\IZ_{20}$| $4\cdot 20$ | $({5\over 20},{2\over 20},{1\over 20},0)$ | 4 | 22
\nr
$\IZ_{60}$| $3\cdot 4\cdot 5 \cdot 60/1\cdot 12\cdot 15\cdot 20$ |
  $({10\over 60},{4 \over 60},{1\over 60},{1\over 60},{1\over 60},
  {1\over 60})$ | 2 | 24
\endtable
}

\appendix{B}{Examples of cyclic orbifolds of the monster module CFT}

The first column is the conjugacy
class of the element in $g\in\IM$. The second column gives the
eigengroup $E(g)$, and the third column gives the shift vector.

\vbox{
\begintable
$g$| E(g)| $\delta_R$   \cr
2A | 2+ | $(\half,\half,\half,\half)$ \nr
2B | 2- |  $(\half,\half,\half,\half)$ \nr
3A | 3+ |  $({2\over 3},{1\over 3}, {1\over 3},0 )$ \nr
3B | 3- |  $({2\over 3},{1\over 3}, {1\over 3},0 )$ \nr
4D | $4\vert 2-$ | $({1\over 4},{1\over 4},{1\over 4},{1\over 4})$ \nr
5A | 5+ |  $({3\over 5}, {1\over 5},0,0)$ \nr
5B | 5- |  $({3\over 5}, {1\over 5},0,0)$ \nr
7A | 7+ | $({3\over 7}, {2\over 7}, {1\over 7},0)$ \nr
7B | 7- | $({3\over 7}, {2\over 7}, {1\over 7},0)$ \nr
11A | 11+ | $({4\over 11}, {2\over 11},{1\over 11},
{1\over 11})$ \nr
13B | 13- | $({5\over 13}, {1\over 13},0,0)$
\endtable
}

\listrefs
\end